\documentclass[twoside,a4paper,11pt]{sea10}
\usepackage{graphicx}
\usepackage{hyperref}
\usepackage{movie15}

\usepackage{natbib}

\newcommand{\mz}{{\small $\mathcal{M}$-Z}}
\newcommand{\Sz}{{\small {\large $\Sigma$}-Z}}

\DeclareRobustCommand{\ion}[2]{%
\relax\ifmmode
\ifx\testbx\f@series
{\mathbf{#1\,\mathsc{#2}}}\else
{\mathrm{#1\,\mathsc{#2}}}\fi
\else\textup{#1\,{\mdseries\textsc{#2}}}%
\fi}
\newcommand{\HII}{\ion{H}{ii}~}

\newcommand{\pasp}{Pub. Astr. Soc. of Pacific}

\newcommand{\apj}{Astrophysical Journal}
\newcommand{\apjl}{Astrophysical Journal Letter}
\newcommand{\mnras}{Monthy Notices of Royal Astronomial Society}
\newcommand{\aap}{A\&A}

\begin{document}
\pagenumbering{arabic}
\pagestyle{myheadings}
\thispagestyle{empty}
{\flushleft\includegraphics[width=\textwidth,bb=58 650 590 680]{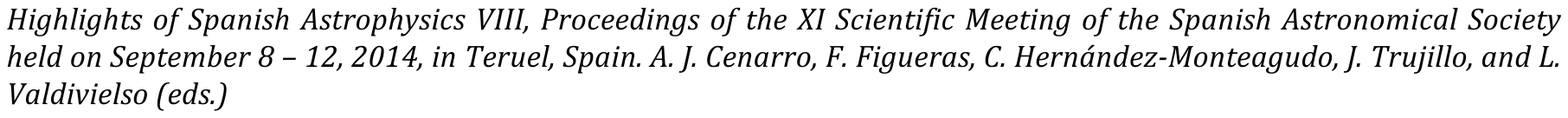}}
\vspace*{0.2cm}
\begin{flushleft}
{\bf {\LARGE
%
The CALIFA survey: Status Report
%
}\\
\vspace*{1cm}
%
S.F.\,S\'anchez$^{1}$,
J.M.\,Vilchez$^{2}$,
C.J.\,Walcher$^{3}$,
R\,Gonzalez Delgado$^{2}$,
A\,Gil de Paz$^{4}$,
P\,S\'anchez Blazquez$^{5}$,
R\,Marino$^{4}$,
and 
The CALIFA collaboration
%
}\\
\vspace*{0.5cm}
%
$^{1}$
Instituto de Astronom\'\i a,Universidad Nacional Auton\'oma de Mexico, A.P. 70-264, 04510, M\'exico,D.F.\\
$^{2}$
Instituto de Astrof\'{\i}sica de Andaluc\'{\i}a (CSIC), Glorieta de la Astronom\'\i a s/n, Aptdo. 3004, E18080-Granada, Spain.\\
$^{3}$
Leibniz-Institut f\"ur Astrophysik Potsdam (AIP), An der Sternwarte 16, D-14482 Potsdam, Germany.\\
$^{4}$
Departamento de Astrof\'{i}sica y CC$.$ de la Atm\'{o}sfera, Facultad de CC$.$ F\'{i}sicas, Universidad Complutense de Madrid, Avda.\,Complutense s/n, 28040 Madrid, Spain.\\
$^{5}$
Departamento de F\'isica Te\'orica, Universidad Aut\'onoma de Madrid, 28049 Madrid, Spain.\\
%
\end{flushleft}
%
\markboth{
CALIFA status report
}{ 
%
S.F.\,S\'anchez et al.
%
}
\thispagestyle{empty}
\vspace*{0.4cm}
\begin{minipage}[l]{0.09\textwidth}
\ 
\end{minipage}
\begin{minipage}[r]{0.9\textwidth}
\vspace{1cm}
\section*{Abstract}{\small
%
We present here a brief summary of the status of the on-going 
CALIFA survey with an emphasis on the results that have been recently
published. In particular, we make a summary of the most relevant results 
found regarding the properties of \HII regions discovered using this survey,
and the evidence uncovered for an inside-out growth of galaxies.
%
\normalsize}
\end{minipage}
%
%
%
\section{Introduction \label{intro}}

The Calar Alto Legacy Integral Field Area (CALIFA) survey
\citep{sanchez12} is an ongoing large project of the Centro
Astron\'omico Hispano-Alem\'an at the Calar Alto observatory to obtain
spatially resolved spectra for 600 local (0.005$<z<$0.03) galaxies by
means of integral field spectroscopy (IFS). CALIFA observations
started in June 2010 with the Potsdam Multi Aperture Spectrograph
(PMAS), mounted to the 3.5 m telescope, utilizing the large
(74$"$$\times$64$"$) hexagonal field-of-view (FoV) offered by the PPak
fiber bundle \citep{verheijen04,kelz06}. PPak was created for the Disk
Mass Survey (Bershady et al. 2010). Each galaxy is observed using two
different setups, an intermediate spectral resolution one (V1200,
$R\sim 1650$), that cover the blue range of the optical wavelength
range (3700-4700\AA), and a low-resolution one (V500, $R\sim 850$, that
covers the first octave of the optical wavelength range
(3750-7500\AA). A diameter-selected sample of 939 galaxies were drawn
from the 7th data release of the Sloan Digital Sky Survey (SDSS) which
is described in \cite{walcher14}. From this mother sample the 600
target galaxies are randomly selected, of which we have currently
observed 517 objects (December 2014), being near to its conclusion.

Combining the techniques of imaging and spectroscopy through optical
IFS provides a more comprehensive view of individual galaxy properties
than any traditional survey. CALIFA-like observations were collected
during the feasibility studies (M\'armol-Queralt\'o et al. 2011; Viironen
et al. 2012) and the PPak IFS Nearby Galaxy Survey (PINGS,
Rosales-Ortega et al. 2010), a predecessor of this survey. First
results based on those datasets already explored their information
content (e.g. Rosales-Ortega et al. 2010;  Rosales-Ortega et
al. 2012). 

Compared with other IFS surveys, CALIFA offers an unique combination
of (i) a sample covering a wide range of morphological types in a wide
range of masses, sampling the Color-Magnitude diagram for M$_g>-$ 18
mag; (ii) a large FoV, that guarantees to cover the entire optical
extension of the galaxies up to 2.5$r_e$ for an 80\% of the sample;
and (iii) an accurate spatial sampling, with a typical spatial
resolution of $\sim$1 kpc for the entire sample, which allows to
optical spatial resolved spectroscopic properties of most relevant
structures in galaxies (spiral arms, bars, buges, Hii regions...). The
penalty for a better spatial sampling of the galaxies is the somehow
limited number of galaxies in the survey, e.g., MaNGA \citep{bundy14}
and SAMI \citep{sami}. In terms of the spectral resolution, while in
the red both survey have better spectral resolution than CALIFA, in the the
blue wavelength range both three have similar resolutions.

As a legacy survey, one of the main goals of the CALIFA collaboration
is to grant public access of the fully reduced datacubes. In November
2012 we deliver our 1st Data Release (Husemann et al. 2013),
comprising 200 datacubes corresponding to 100
objects \footnote{http://califa.caha.es/DR1/}. After almost two years,
and a major improvement in the data reduction, we present our 2nd Data Release
(Garcia Benito et al., 2014), comprising 400 datacubes corresponding to 200
objects \footnote{http://califa.caha.es/DR2/}, the 1st of October 2014.

\section{CALIFA: Main Science Results}

The data products that can be derived from the IFU datasets obtained by
the CALIFA survey comprise information on the stellar populations,
ionized gas, mass distribution and stellar and gas kinematics. Similar
data products are derived for any of the indicated projects: Atlas3D,
MaNGA or SAMI. In summary, they conform a panoramic view of the
spatial resolved spectroscopic prorperties of these galaxies

Different science goals have been already addressed using this
information: (i) New techniques has been developed to understand the
spatially resolved star formation histories (SFH) of galaxies (Cid
Fernandes et al., 2013, 2014). We found the solid evidence that
mass-assembly in the typical galaxies happens from inside-out \citep{perez13}. The SFH and chemical enrichment of bulges and early-type galaxies are
fundamentally related to the total stellar mass, while for disk
galaxies it is more related to the local stellar mass density
\citet{rosa13,rosa14}; negative age gradients indicate that the
quenching is progressing outward in massive galaxies \citep{rosa13},
and age and metallicity gradients suggest that galaxy bars have not alter signicantly
the SFH of spirals \citep{patri14}; (ii) We explore the origin
of the low intensity, LINER-like, ionized gas in galaxies. These
regions are clearly not related to star-formation activity, or
to AGN activity. They are most probably relatd to post-AGB ionization
in many cases \citep{papa13}; (ii) We explore the aperture and resolution effects on the
data. CALIFA provides a unique tool to understand the aperture and
resolution effects in larger single-fiber (like SDSS) and IFS surveys
(like MaNGA, SAMI). We explored the effects of the dilution of the
signal in different gas and stellar population properties (Mast et
al., 2014), and proposed an new empirical aperture correction for the
SDSS data \citep{iglesias13}; (iv) CALIFA is the first
IFU survey that allows gas and stellar kinematic studies for all
morphologies with enough spectroscopic resolution to study (a) the
kinematics of the ionized gas \citep{bego14}, (b) the
effects of bars in the kinematics of galaxies \citep{jkbb14}; (c) the effects of the intraction stage on the kinematic
signatures (Barrera-Ballesteros et al., submitted), (d) measure the
Bar Pattern Speeds in late-type galaxies (Aguerri et al., submitted),
(iv) extend the measurements of the angular momentum of galaxies to
previously unexplored ranges of morphology and ellipticity
(Falc\'on-Barroso et al., in prep.); and (v) finally we explore in
detail the effects of galaxy interaction in the enhancement of
star-formation rate and the ignition of galactic outflows \citep{wild14}. The results based focused on the analysis of the \HII
regions will be discussed in the next Section.

\section{Results of our studies of the \HII regions}\label{high}

The program to derive the properties of the \HII regions
was initiated based on the data from the PINGS survey
\citet{rosales-ortega10}. This survey acquired IFS mosaic data for a dozen
of medium size nearby galaxies. In \citep{sanchez11} and
\citet{rosales11} we studied in detail the ionized gas and \ion{H}{ii}
regions of the largest galaxy in the sample (NGC\,628). The main
results of this studies are included in the contribution by
Rosales-Ortega in the current edition. We then continued the
acquisition of IFS data for a larger sample of visually classified
face-on spiral galaxies \citep{marmol-queralto11}, as part of the
feasibility studies for the CALIFA survey \citep{sanchez12}. The
spatially resolved properties of a typical galaxy in this sample,
UGC9837, were presented by\citep{viir12}.

In \citep{sanchez12b} we presented a new method to detect, segregate
and extract the main spectroscopic properties of \ion{H}{ii} regions
from IFS data
(\textsc{HIIexplorer}). A preliminar catalog of  $\sim$2600 \ion{H}{ii} regions and aggregations
extracted from 38 face-on spiral galaxies compiled from the PINGS and
CALIFA feasibility studies was presented. We found
a new local scaling relation between the stellar mass density and
oxygen abundance, the so-called \Sz\ relation \citep{rosales12}.

The same catalog allows us explore the galactocentric radial gradient of the
oxygen abundance \citep{sanchez12b}. We confirmed that up
to $\sim$2 disk effective radius there is a negative gradient of the
oxygen abundance in all the analyzed spiral galaxies. The
gradient presents a very similar slope for all the galaxies ($\sim
-0.12$ dex/$r_e$), when the radial distances are measured in units of
the disk effective radii. Beyond $\sim$2 disk effective radii our data show evidence of a flattening in the abundance, consistent with several
other spectroscopic explorations, based mostly on a few objects
\citep[e.g.][]{bresolin09}.

\begin{figure}
\centering
\includegraphics[width=7.6cm]{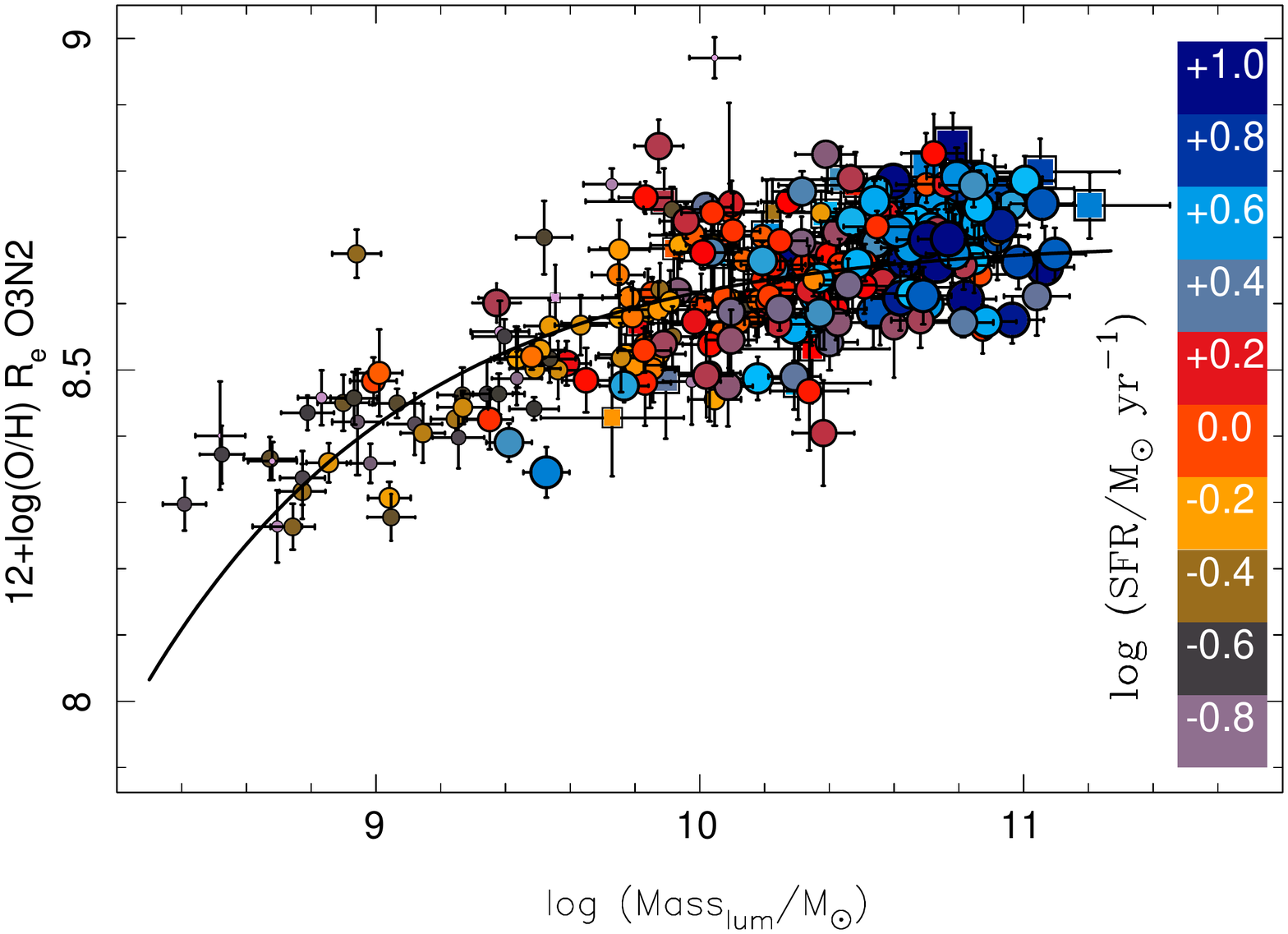}
\includegraphics[width=7.6cm]{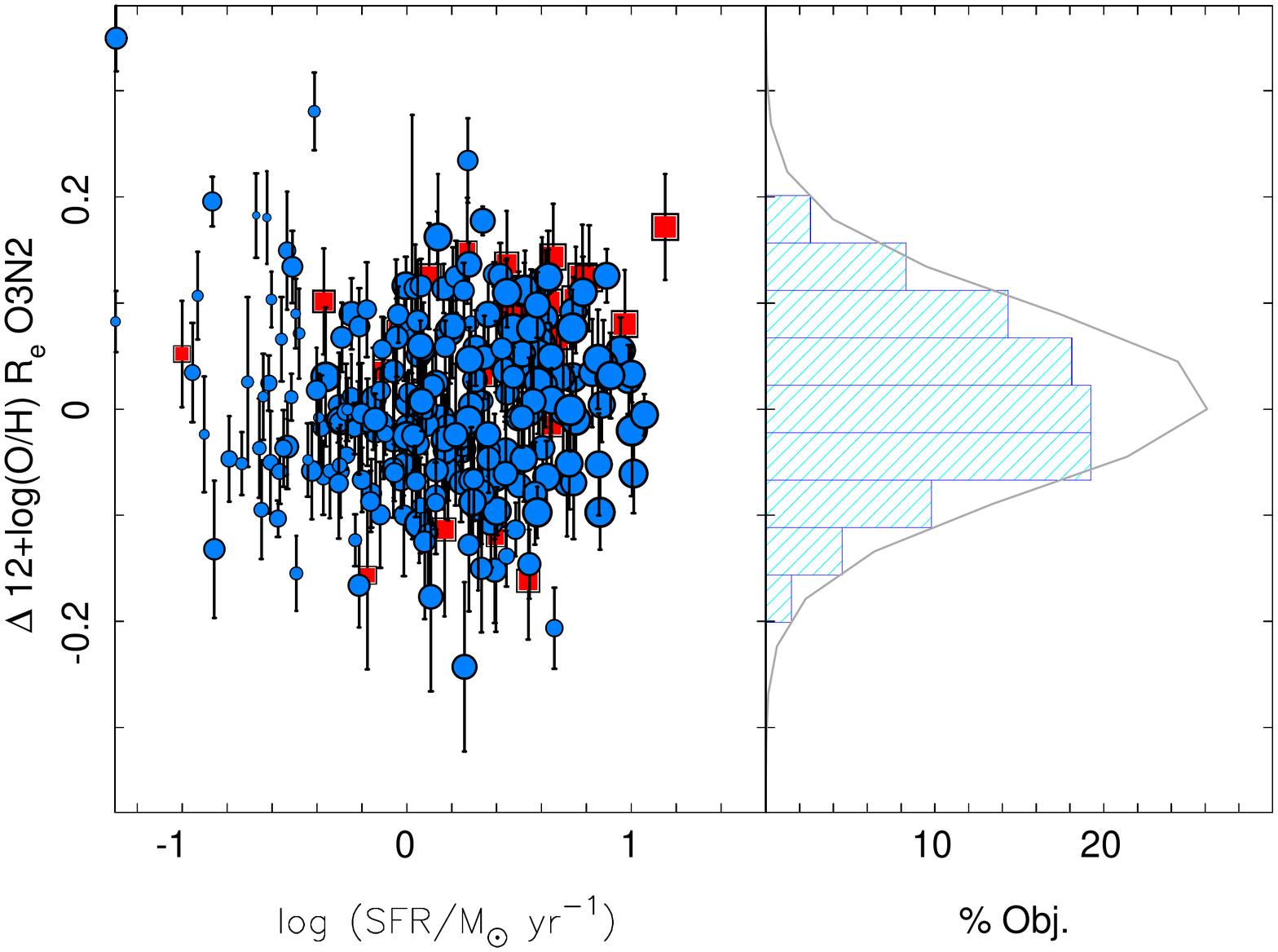}
\caption{\label{fig:MZ}. {\it Left panel:} Distribution of the oxygen abundances at the effective radii as a function of the integrated stellar masses for the CALIFA galaxies (236, circles), together with those from the CALIFA feasibility studies (31, squares). {\it Right Panel:} Distribution of the differential oxygen abundances with respect to the solid-line shown in the left-panel (i.e., the dependence on the stellar mass), as a function of the integrated SFR for the CALIFA galaxies.}
\end{figure}

In \citep{sanchez13} we presented the first results based on the
catalog of \HII\ regions extracted from a enlarged sample of
galaxies ($\sim$100). We studied the dependence of the \mz\ relation
with the star formation rate. We found
that no secondary relation different than the one induced by the well
known relation between the star formation and the mass, contrary to
what was claimed other authors \citep{lara10a,mann10}, based on single
aperture spectroscopic data (SDSS). Although the reason for the
discrepancy is still not clear, we postulate that simple aperture
bias, like the one present in previous datasets, may induce the
reported secondary relation. Figure \ref{fig:MZ} presents an updated
version of these results, including the last list of analyzed
galaxies, until July 2014 (236 galaxies from the CALIFA sample
together with 31 galaxies from the CALIFA-pilot studies). The left
panel shows the \mz\ relation found for these galaxies, with color
code indicating the integrated SFR for each galaxy. It is appreciated
that the stronger gradient in SFR is along the stellar mass, as
expected for star-forming galaxies. Once subtracted the best fitted
function to the \mz\ relation, the residual of the abundance do not
present any evident secondary relation with the SFR (as it is seen in
the right panel). Thus, the results presented in \citep{sanchez13} are
confirmed with a sample of galaxies enlarged by almost a factor two.

We also confirmed the local \Sz\ relation unveiled by
\citet{rosales11}, with a larger statistical sample of \HII regions
($\sim$5000). This nebular gas \Sz\ relation is flatter than
the one derived for the average stellar populations \citep{rosa14}, but
both of them agree for the younger stars, as expected if the most
recent stars are born from the chmical enriched ISM.  In \citep{sanchez14}, we confirmed that the abundance
gradients present a common slope up to $\sim$2 effective radii, with a
distribution compatible with being produced by random fluctuations,
for all galaxies when normalized to the disk effective radius of
$\alpha_{O/H}=-$0.1 dex/$r_e$.  Finally, in \citep{sanchez14b}, we
found evidence that \HII regions keep a memory of their past, by
analysing the correspondance between the properties of these ionized
regiones with that of their underlying stellar populations.

\section{Conclusions}

In summary the results from the CALIFA survey present a coherent picture of
the mass-growth and chemical enrichment of galaxies. All the results
indicate that the bulk of the galaxy population presents an
inside-out growth (at the mass range covered by the survey),
with a chemical enrichment dominated by local processes, and
limited effects by processes like outflows or radial mixing.


%
\end{document}